\documentclass[10pt]{article} % For LaTeX2e

\usepackage[accepted]{rlj} % Should be uncommented for the camera-ready
\usepackage{amssymb}            % Defines common symbols like \mathbb R
\usepackage{mathtools}          % Extends amsmath, providing common math tools
\usepackage{mathrsfs}           % Enables \mathscr, which can work in cases that \mathcal does not
\usepackage{graphicx}           % For including images
\usepackage{subcaption}         % Allows for the use of subfigures and subcaptions
\usepackage[space]{grffile}     % For spaces in image names
\usepackage{url}                % For displaying URLs
\usepackage{lipsum}             % For placeholder text

\usepackage{algorithm}
\usepackage{algpseudocode}
\usepackage{multirow}
\usepackage{flafter}
\title{Learning Bilateral Team Formation in Cooperative Multi-Agent Reinforcement Learning}

\setrunningtitle{Learning Bilateral Team Formation in Cooperative MARL}

\author{Koorosh Moslemi\textsuperscript{1,$\dagger$}, Chi-Guhn Lee\textsuperscript{1}}

\emails{koorosh.moslemi@mail.utoronto.ca, chiguhn.lee@utoronto.ca}

\affiliations{
$^{1}$\textbf{University of Toronto, Canada}
\par % If including additional comments like below, use \par to add some whitespace. 
$^\dagger$ Corresponding author
}

\contribution{
    We study the team formation problem in MARL as a bilateral matching between two disjoint sets of agents. We empirically show a stable matching algorithm results in policies that generalize better to unseen agent compositions compared to those derived from an unstable matching.
    }
    {
    The matching market problem involves matching two disjoint sets based on their preferences. A key property in this domain is \textit{stability} \citep{gale1962college}, which ensures that no agent has an incentive to deviate by seeking alternative matches. Within this domain, the problem of learning preferences has primarily been explored in a bandit framework. While previous research has examined aspects of stable matchings such as centralized versus decentralized \citep{liu2020competing}, one-sided versus two-sided preference learning \citep{zhangdecentralized}, and regret analysis in bandit settings, our work shifts the focus to MARL. Specifically, we investigate the performance of stable and unstable matchings in MARL, providing new insights into their comparative effectiveness.
    }

\contribution{
     We propose an attention-based value decomposition method for team formation, introducing targeted modifications to standard value and mixing networks. This framework enables learning bilateral team formation for a dynamic population of agents.
    }
    {
    The core contribution of our work lies in interpreting attention scores among agents as preferences and leveraging matching algorithms to process these learned preferences. While prior studies, such as GoMARL \citep{zang2024automatic} and VAST \citep{phan2021vast}, investigate the influence of subgroups on value decomposition within a fixed agent population, our framework relaxes this assumption by only constraining the maximum number of agents. Alternative approaches, such as COPA \citep{liu2021coach}, require pre-defined teams during training to bypass learning the team formation process. Others adopt unilateral teaming methods, such as SOG \citep{shao2022self}, which relies on random selection for one side of the matching process.
    }

 % Your keywords

\summary{Team formation and the dynamics of team-based learning have drawn significant interest in the context of Multi-Agent Reinforcement Learning (MARL). However, existing studies primarily focus on unilateral groupings, predefined teams, or fixed-population settings, leaving the effects of algorithmic bilateral grouping choices in dynamic populations underexplored. To address this gap, we introduce a framework for learning two-sided team formation in dynamic multi-agent systems. Through this study, we gain insight into what algorithmic properties in bilateral team formation influence policy performance and generalization. We validate our approach using widely adopted multi-agent scenarios, demonstrating competitive performance and improved generalization in most scenarios.
}

\begin{document}

\makeCover  % Create the cover page
\maketitle  % Make the title section

\begin{abstract}
Team formation and the dynamics of team-based learning have drawn significant interest in the context of Multi-Agent Reinforcement Learning (MARL). However, existing studies primarily focus on unilateral groupings, predefined teams, or fixed-population settings, leaving the effects of algorithmic bilateral grouping choices in dynamic populations underexplored. To address this gap, we introduce a framework for learning two-sided team formation in dynamic multi-agent systems. Through this study, we gain insight into what algorithmic properties in bilateral team formation influence policy performance and generalization. We validate our approach using widely adopted multi-agent scenarios, demonstrating competitive performance and improved generalization in most scenarios.
\end{abstract}

%%%%%%%%%%%%%%%%%%%%%%%%%%%%%%%%%%%%%%%%%%%%%%%%%%%%%%%%%%%%%%%%
%% Section: Submission of papers to RLJ/RLC
%%%%%%%%%%%%%%%%%%%%%%%%%%%%%%%%%%%%%%%%%%%%%%%%%%%%%%%%%%%%%%%%

\section{Introduction}

In recent years, Multi-Agent Reinforcement Learning (MARL) has emerged as a powerful framework for tackling complex decision-making problems that require coordinated actions across multiple autonomous agents. Such systems find application in various domains, from traffic control \citep{zhang2024field} and autonomous vehicles \citep{lee2025traffic} to robotics \citep{krnjaic2024scalable}. Within this framework, team formation has gained particular attention, as the ability to form, adapt, and coordinate effective teams of agents creates additional possibilities for incorporating valuable group-level learning cues.

In the realm of team formation, value decomposition methods have gained traction in recent years. These methods typically break down the centralized action-value function into individual value functions, each conditioned on the state and actions of a single agent. However, recent works such as VAST \citep{phan2021vast} and GoMARL \citep{zang2024automatic} have highlighted performance bottlenecks in such flat decompositions. To address this, they propose a group-wise factorization approach. Despite these advancements, there is no established consensus on how to form these groups. Existing methods have explored various strategies, including leveraging clustering \citep{phan2021vast}, heuristic-based approaches (e.g., initially placing all agents in a single group and gradually reallocating them based on predefined rules) \citep{zang2024automatic}, one-sided matching between two disjoint agent subsets \citep{shao2022self}, and implicit grouping by learning to allocate tasks to agents \citep{wang2020rode} in hierarchical reinforcement learning literature.

In a different research domain, the matching market problem—which involves matching two disjoint sets based on their preferences—has been extensively studied in applications such as college admissions \citep{gale1962college}, labor markets \citep{roth1999redesign}, and medical residency programs \citep{roth1984evolution}. A key property in this domain is stability, which ensures that no agent has an incentive to deviate by seeking alternative matches. Since agents' preferences may be uncertain or unknown depending on the application, significant research has focused on learning these preferences, often within a bandit framework. However, to the best of our knowledge, the role of stability of matching and two-sided preference learning has not been examined in the context of team formation in MARL. 

In MARL, nonstationarity poses a fundamental challenge—agents continuously improve their skills over time while the environment itself changes. Learning preferences in such a dynamic setting is significantly more complex than in traditional multi-armed bandit problems, where the reward distribution of each arm remains stationary. A key distinction in MARL is the interdependence between matching and preference learning: the way agents are matched influences the environment dynamics, which in turn affects preference learning, while preferences themselves shape the outcomes of matching. This interplay raises a crucial question: what properties of a matching mechanism facilitate effective learning? To illustrate, consider a heterogeneous team of ground and aerial robots deployed in a search-and-rescue scenario. Ground robots must learn to navigate rough terrain, remove obstacles, and transport supplies, while aerial robots focus on mapping, scouting, and relaying communication. For effective operation, these robots must form cooperative ground-aerial pairs while simultaneously learning their preferences—such as camera resolution, and low-latency communication.
Since both ground and aerial robots are continuously learning, they may develop conflicting perspectives on optimal pairings, despite being trained for a common objective. A matching mechanism aggregates differing views, influencing the overall learning outcome.
This leads to an important inquiry: does a property like stability in matching (i.e., reducing the likelihood of agents frequently switching partners) improve resulting policies? To address this question, we make the following contributions:

\begin{itemize}
 \item We study the team formation problem in MARL as a bilateral matching between two disjoint sets of agents. We empirically show a stable matching algorithm results in policies that generalize better to unseen agent compositions compared to those derived from an unstable matching.
 \item We propose an attention-based value decomposition method for team formation, introducing targeted modifications to standard value and mixing networks. This framework enables learning bilateral team formation for a dynamic population of agents.
\end{itemize}

\section{Related Work}

\paragraph{Team Formation} \citet{zang2024automatic} introduced GoMARL, a grouping method designed for a fixed population of agents. In this method, all agents are initially in one group, and they're gradually reassigned to other groups based on a heuristic that evaluates their fitness to the current group. By learning an adaptive group structure, GoMARL learns the state-action value function for groups of agents, addressing performance bottlenecks in flat decomposition methods that rely on individual value functions. Similarly, our method employs group-wise value decomposition. However, unlike GoMARL, the algorithms we explore are not tied to a specific learning paradigm (i.e., they can be applied to both policy gradient and value-based methods). Additionally, our approach accommodates a variable number of agents at execution, provided that the maximum number is predetermined during training. 
\citet{liu2021coach} proposed COPA, which introduced a coach agent with global information to periodically broadcast strategies to a dynamic number of agents. COPA departs from the common decentralized execution (DE) assumption, arguing that it is too restrictive for complex tasks. Our work similarly challenges this assumption by enforcing structured group formation during execution. However, unlike COPA which relies on a predefined set of teams during training, we learn inter-agent preferences that lead to matching agents and forming teams.
SOG \citep{shao2022self} explored the idea of matching two disjoint sets of agents—conductors and non-conductors. Conductor agents, either randomly selected or elected, send group invitations to non-conductor agents, who then choose among multiple invitations at random. In contrast, our work focuses on bilateral team formation, where matching is driven by mutual preferences rather than unilateral selection. Within the DE paradigm, CollaQ \citep{zhang2020multi} and MIPI \citep{ye2024mutual} addressed generalization challenges in dynamic team formation.  Similarly, our work aims to develop a method that results in moderately close performance between training and execution as the number of agents increases at execution. However, we specifically examine how the properties of matching algorithms, such as stability versus instability, affect team formation in a centralized execution setting.

\paragraph{Matching and Learning Preferences}
The problem of learning preferences in matching markets is often studied within a multi-armed bandit framework, where agents have uncertain \citep{aziz2020stable} or unknown preferences over arms, while the arms' preferences over agents are sometimes assumed to be known  \citep{liu2020competing,hosseini2024putting}. The primary objective in this literature is to design algorithms that minimize regret for each agent. \citet{liu2020competing} formalize both centralized and decentralized bandit learning for matching markets. More recently, \citet{zhangdecentralized} extend this framework to settings where preferences on both sides are unknown. \citet{wang2022bandit} and \citet{kong2024improved} further generalize the problem from one-to-one to many-to-one matching. Similarly, we investigate a many-to-one matching problem where preferences on both sides are unknown. However, in our case, matching occurs among agents collaboratively optimizing the expected discounted sum of rewards in a MARL setting.

\section{Background}
\label{apndx:team_formation_background}
\subsection{Entity-wise Dec-POMDP}

We study the \textit{decentralized partially observable Markov decision process} (Dec-POMDP) \citep{oliehoek2016concise} with entities \citep{schroeder2019multi} described as $
(\mathbf{S}, \mathbf{U}, \mathbf{O}, P, r, \mathcal{E}, \mathcal{A}, \mu)$. $\mathcal{E}$ represents entities in the environment, either agents or non-agents (i.e., $\mathcal{A} \subseteq \mathcal{E}$). Unlike the set of agents $\mathcal{A}$, non-agent entities cannot be controlled by learning policies (e.g., obstacles, enemies with fixed behavior). $\mathbf{s} \in \mathbf{S}$ denotes the global state containing state representation for entities $s^e$ (i.e., $\mathbf{s} = \left\{s^e \mid e \in \mathcal{E}\right\} \in \mathbf{S}$). Each agent's partial observability $o^a \in \mathbf{O}$ is defined as $\left\{s^e \mid \mu\left(s^a, s^e\right)=1, e \in \mathcal{E}, a \in \mathcal{A} \right\}$ where $s^a$ is the state representation for agent $a \in \mathcal{A}$, and $\mu\left(s^a, s^e\right) \in \left\{ 0,1 \right\}$ is the binary observability mask. In other words,  $\mu\left(s^a, s^e\right)=1$ means agent $a \in \mathcal{A}$ observes entity $e \in \mathcal{E}$. $\mathbf{U}$ is the set of joint actions of all agents. Additionally, $P$ and $r$ denote the state transition and reward functions, respectively.

To deal with a dynamic population of agents, a multi-head attention module, denoted as $\text{MHA}(\mathcal{A},\textit{eFF}(\mathbf{X}^\mathcal{E}),\mathbf{M})$, integrates information not blocked by masks.
$\mathbf{X}^\mathcal{E} \in \mathbb{R}^{|\mathcal{E}|\times d}$ is the entity representation matrix where $d$ is the dimensionality of the input. $\textit{eFF}(.)$ denotes an entity-wise feedforward layer, which is a standard fully connected layer that applies identical transformations to all input entities.
$\mathbf{M} \in \mathbb{R}^{|\mathcal{A}|\times |\mathcal{E}|}$ is the mask matrix which specifies which entities can be queried by agents. For more details on the attention module please refer to \citep{iqbalsupplementary}. Additionally, we use the subscript $t$ to denote the timestep and define the discount factor as $\gamma \in [0,1)$.

\subsection{Value Decomposition}
% Qmix + VDN
Value decomposition methods \citep{sunehag2017value} decompose the centralized action-value function $Q^{\text{tot}}(\mathbf{s},\mathbf{u})$ into simpler value functions (e.g., individual value functions that are conditioned on states and actions of one agent). This decomposition addresses the exponential growth of the joint action space with the number of agents and enables decentralized action selection. The centralized action-value function is defined as:
\begin{equation*}
    Q^{\text{tot}}(\mathbf{s},\mathbf{u}) := r(\mathbf{s}, \mathbf{u})+\gamma \mathbb{E}\left[\max Q^{\mathrm{\text{tot}}}\left(\mathbf{s}^{\prime}, \cdot\right) \mid s^{\prime} \sim P(\cdot \mid \mathbf{s}, \mathbf{u})\right].
\end{equation*}
QMIX \citep{rashid2018qmix} uses a monotonic mixing function $f_{\textit{mix}}$ that combines individual utilities to approximate $Q^{\text{tot}}$. The monotonicity of $f_{\textit{mix}}$ ensures an increase in the value of any agent $Q^a$ for its action $u^a$ leads to an increase in $Q^{\text{tot}}$ for joint actions $\mathbf{u} \in \mathbf{U}$. Consequently, composing greedy individual actions with respect to individual value functions results in a greedy joint action with respect to $Q^{\text{tot}}$. In other words, $f_{\textit{mix}}$ is used as follows:
\begin{align*}
 Q_\theta^{\mathrm{tot}}\left(\boldsymbol{\tau}_t, \mathbf{u}_t\right) &\approx f_{\textit{mix}}\left(Q^1\left(\tau_t^1, u_t^1 ; \theta_Q\right), \ldots, Q^{|\mathcal{A}|}\left(\tau_t^{|\mathcal{A}|}, u_t^{|\mathcal{A}|} ; \theta_Q\right) ; \theta\right), \\
\theta &= h(\boldsymbol{s}_t; \theta_h).
\end{align*}
where $Q_\theta^{\mathrm{tot}}\left(\boldsymbol{\tau}_t, \mathbf{u}_t\right) \approx Q^{\mathrm{tot}}\left(\mathbf{s}_t, \mathbf{u}_t\right)$ is the parametrized centralized value function. This function uses trajectories $\boldsymbol{\tau}_t:=\left\{\tau_t^a\right\}_{a \in \mathcal{A}}$, where $\tau_t^a := \left(o_0^a, u_0^a, \ldots, o_t^a\right)$ denotes history of an agent's actions and observations, as a proxy for global state.
$h$ is a hyper-network \citep{ha2022hypernetworks} conditioned on the global state and parametrized by $\theta_h$. In practice, the monotonicity assumption is ensured by constraining $\theta$ to be positive using techniques such as applying the absolute value function or softmax normalization. To learn the Q-function, the following objective is minimized:
\begin{align}
\mathcal{L}_{Q}(\theta) & :=\mathbb{E}\left[\left(y_t^{\mathrm{tot}}-Q_\theta^{\mathrm{tot}}\left(\boldsymbol{\tau}_t, \mathbf{u}_t\right)\right)^2 \mid\left(\boldsymbol{\tau}_t, \mathbf{u}_t, r_t, \boldsymbol{\tau}_{t+1}\right) \sim \mathcal{D}\right] \label{eq1}, \\
y_t^{\mathrm{tot}} & :=r_t+\gamma Q_{\bar{\theta}}^{\mathrm{tot}}\left(\boldsymbol{\tau}_{t+1}, \arg \max Q_\theta^{\mathrm{tot}}\left(\boldsymbol{\tau}_{t+1}, \cdot\right)\right) .\nonumber 
\end{align}

where $Q_{\bar{\theta}}^{\mathrm{tot}}$ is a target network \citep{mnih2015human} with weights $\bar{\theta}$ that are copied from $\theta$ periodically to stabilize the regression target $y_t^{tot}$. $\mathcal{D}$ is a replay buffer \citep{lin1992self} containing experiences of all agents collected by a policy for exploration.

\section{Method}

\label{apndx:team_formation_details}

In this section, we detail the modifications made to agents’ utility (a.k.a., value)
and mixing networks within a value decomposition framework. We build on REFIL \citep{iqbal2021randomized}, as it already incorporates an attention mechanism capable of handling a dynamic number of agents. In short, these modifications include changes to the architecture of the models (e.g., adding encoder-decoder networks to enforce group structures) and training objectives (e.g. encouraging similarity within groups and diversity among groups).

\begin{figure}
    \centering
    \subfloat[]{{\includegraphics[width=0.45\linewidth]{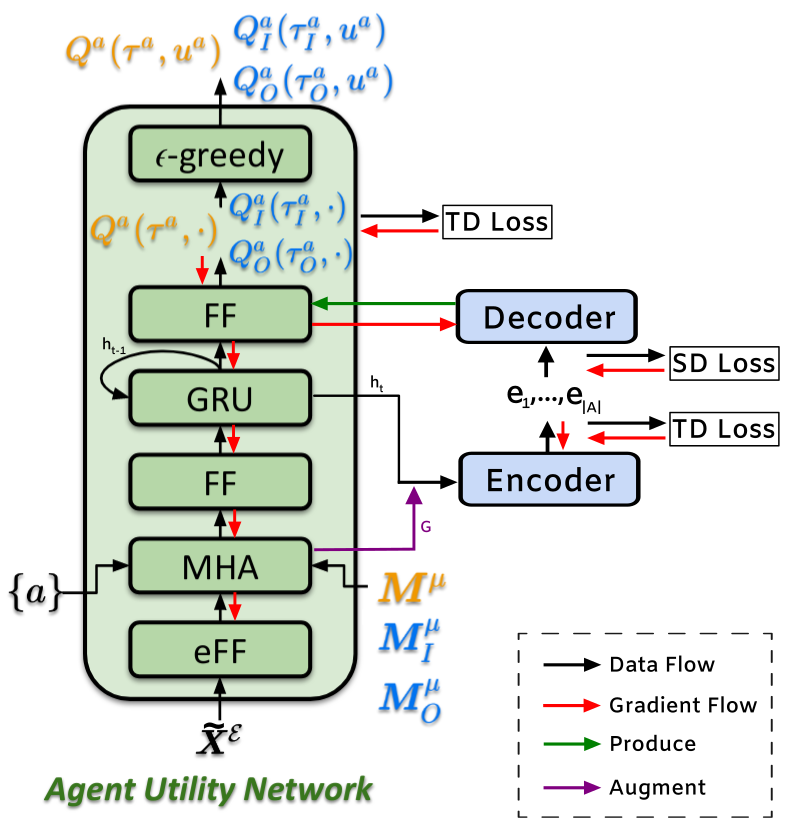}}\label{fig:utility_net}}
    \subfloat[]{{\includegraphics[width=0.45\linewidth]{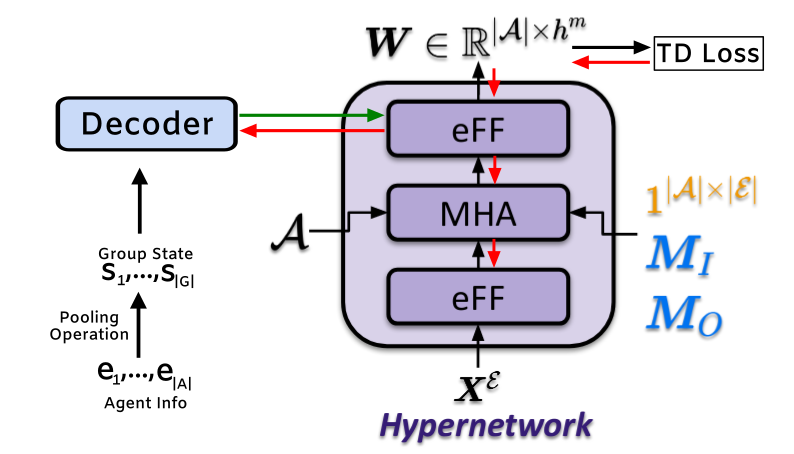}}\label{fig:hyper_net}}
    \caption{Modifications to agent utility network and hypernetwork. (a) An encoder-decoder structure is integrated to incorporate group-related information into the action selection process. The type of utility used in Equations \ref{eq2} and \ref{eq3} (i.e., $Q^a, Q_I^a, Q_O^a$) depends on the input mask of the multi-head attention (MHA) module as color-coded in Figure \ref{fig:utility_net} (b) A group-aware hypernetwork is employed to generate the weights for the mixing network $f_{\textit{mix}}$. The decoder used here is different from Figure \ref{fig:utility_net} as they generate weights for different layers. However, the input embeddings $ e_1, \cdots, e_{|\mathcal{A}|}$ passed to both decoders are the same. }
\end{figure}

\subsection{Agent Utility Network} \label{utility_net}
As illustrated in Figure \ref{fig:utility_net}, this network integrates information across entities with an attention module $\text{MHA}(\mathcal{A},\textit{eFF}(\mathbf{\Tilde{X}}^\mathcal{E}),(\mathbf{M}^{\mu},\mathbf{M}_O^{\mu},\mathbf{M}_I^{\mu}))$\footnote{Technically, $\mathbf{\Tilde{X}}^\mathcal{E}$ is repeated to adjust for the three masks. However, we abuse the notation in writing.}. 
$\mathbf{M}_O^{\mu}$ and $\mathbf{M}_I^{\mu}$ are random binary attention masks of size $|\mathcal{A}| \times |\mathcal{E}|$ restricted by the partial observability mask $\mathbf{M}^{\mu}$. These masks enable counterfactual reasoning by introducing randomly created complementary subsets of entities denoted by $I$ and $O$. $\mathbf{\Tilde{X}}^\mathcal{E}$ of size $|\mathcal{E}| \times (d+3)$ is our augmented version of $\mathbf{X}^\mathcal{E}$ with a one-hot vector specifying the type of the entity (i.e., leader, follower, or non-agent entity). We augment the entity representation matrix $\mathbf{X}^\mathcal{E}$ by partitioning the set of agents $\mathcal{A}$ into the set of leader agents (i.e., denoted by $\mathcal{L}$) and follower agents (i.e., denoted by $\mathcal{F}$) as follows:
\[
\mathcal{A} = \mathcal{L} \cup \mathcal{F}, \; \mathcal{L} \cap \mathcal{F} = \emptyset, \; |\mathcal{L}| \leq |\mathcal{F}|.
\]
Our goal is to learn matching between agents in $\mathcal{F}$ and agents in $\mathcal{L}$. To this end, we use attention weights returned by $\text{MHA}(\mathcal{A}, \textit{eFF}(\mathbf{\Tilde{X}}^\mathcal{E}),\mathbf{1}^{|\mathcal{A}| \times |\mathcal{E}|})$ as preferences among agents for team formation. 

Different choices for preference-based matching are discussed in Section \ref{matching_section}. From now on, we assume the groups are denoted with set $\mathcal{G}$.

Inspired by the utility network architecture proposed by \citet{zang2024automatic}, we add an encoder-decoder module to the agent utility network. As illustrated in Figure \ref{fig:utility_net}, the group-aware encoder $f_e(.;\theta_e)$ embeds agents' hidden states such that embeddings corresponding to agents within a group are similar while embeddings corresponding to agents from distinct groups are different. Details for training the encoder $f_e(.;\theta_e)$ are in Section \ref{training_section}. The decoder is a linear transformation that generates parameters of the last feedforward layer in the agent's utility network.
\textbf{These changes have three key benefits: 1) enabling learning of bilateral team formation for a dynamic population of agents; 2) comparing the effect of algorithmic choices for matching on performance and generalization; 3) controlling the composition of new agents (i.e., adding new agents as followers or leaders) in unseen scenarios.}

\subsection{Hypernetwork} \label{hypernet_section}

Similar to the agent's utility network, $\mathbf{X}^\mathcal{E}$ is fed into an entity-wise feedforward layer followed by an attention module $\text{MHA}(\mathcal{A},\textit{eFF}(\mathbf{X}^\mathcal{E}),(\mathbf{1}^{|\mathcal{A}| \times |\mathcal{E}|},\mathbf{M}_O,\mathbf{M}_I))$. Note that all masks ignore partial observability (i.e., replacing $\mathbf{M}^{\mu}$ with an all-ones matrix $\mathbf{1}^{|\mathcal{A}| \times |\mathcal{E}|}$ and not restricting $\mathbf{M}_{O}^{\mu},\mathbf{M}_{I}^{\mu}$  with $\mathbf{M}^{\mu}$) as the hypernetwork is used during centralized training.
As shown in Figure \ref{fig:hyper_net}, embeddings of agents' hidden states generated by the encoder $f_e(.;\theta_e)$ in the agent's utility network are passed through a group-aware pooling operation that generates group-wise states. These states are then fed into a decoder that generates the parameters for the last feedforward layer in the hypernetwork. \textbf{This modification enables factorization of joint action-values by considering group information.}

\subsection{Matching Algorithms} \label{matching_section}
The algorithms we consider in this section require: 1) An inter-agent preference matrix of size $|\mathcal{A}| \times |\mathcal{A}|$ where each row quantifies an agent's preferences on other agents; 2) a prior on the number of leader agents and the size of each team. The inter-agent preference matrix is learned as part of our learning framework by the attention module. Moreover, we explain our choices for the priors in Section \ref{experimental_setup}. The algorithms are described as follows:

\paragraph{Order Oriented Matching (OOM)}

Algorithm \ref{algo:stable_matching} only considers the order of the preferences using a \textit{deferred acceptance} (DA) mechanism and is also known as the many-to-one variant of the stable matching algorithm \citep{gale1962college}. The DA mechanism ensures that the matching outcome is stable, meaning no two agents would prefer to be paired with each other over their current assignments. The algorithm begins with leaders proposing to their top-choice followers until they reach their team size limit. Meanwhile, each follower tentatively accepts the best-received proposal, rejecting any inferior ones. Rejected leaders then propose their next choice. This process repeats until all the leaders have their teams.

\paragraph{Score Oriented Matching (SOM)}
In this algorithm, scores computed by the attention module influence the matching outcome. The core idea is that followers iterate through leaders to find the best free leader based on the mutual score, as shown in Algorithm \ref{alg:best_free_leader}. Unlike OOM, this algorithm does not employ the DA mechanism, making it unstable.

\subsection{Training} \label{training_section}
% introduce sd and td loss
In addition to $\mathcal{L}_{Q}$ in Equation \ref{eq1}, an auxiliary loss $\mathcal{L}_{\textit{aux}}$ is used to factorize Q-function with $2|\mathcal{A}|$ factors (i.e., $Q_I^a$ and $Q_O^a$ illustrated in Figure \ref{fig:utility_net}) by replacing $Q^{\textit{tot}}$ with $Q^{tot}_{\textit{aux}}$:
\begin{align}
Q^{\mathrm{tot}} &=f_{\textit{mix}}\left(Q^1, \ldots, Q^{|\mathcal{A}|} ; h\left(\mathbf{s} ; \theta_h, \mathbf{1}^{|\mathcal{A}| \times |\mathcal{E}|}, \boldsymbol{E}\right)\right), \label{eq2} \\
&\approx Q_{\mathrm{aux}}^{\mathrm{tot}}=f_{\textit{mix}}\left(Q_I^1, \ldots, Q_I^{|\mathcal{A}|}, Q_O^1, \ldots, Q_O^{|\mathcal{A}|}\right. ;
\left.h\left(\mathbf{s} ; \theta_h, \boldsymbol{M}_I, \boldsymbol{E}\right), h\left(\mathbf{s} ; \theta_h, \boldsymbol{M}_O, \boldsymbol{E}\right)\right). \label{eq3}
\end{align}
In the above equations, $h$ denotes the hypernetwork explained in Section \ref{hypernet_section} and $\boldsymbol{E} = \{ e_1, \cdots, e_{|\mathcal{A}|} \}$ is the set of embeddings generated by the encoder $f_e(.;\theta_e)$ introduced in Section \ref{utility_net}. We train the embeddings using the following loss function similar to \citep{zang2024automatic}:
\begin{align} \label{eq4}
& \mathcal{L}_{S D}\left(\theta_e\right)=\mathbb{E}_{\mathcal{B}}\left(\sum_{i \neq j} I(i, j) \cdot \operatorname{cosine}\left(f_e\left(h^i ; \theta_e\right), f_e\left(h^j ; \theta_e\right)\right)\right), \\
& \text { where } I(i, j)= \begin{cases}-1, & a_i, a_j \in g_k, g_k \in \mathcal{G} . \\
1, & a_i \in g_k, a_j \in g_l, k \neq l, g_k,g_l \in \mathcal{G} .
\end{cases} \nonumber
\end{align}

In the above equation, $h^i$ and $h^j$ denote hidden states of agents $a_i$ and $a_j$ respectively. This loss ensures the hidden state embeddings of agents within the same group $g_k$ to be similar. At the same time, it prevents all agents from being alike by encouraging diversity between agents $ a_i \in g_k, a_j \in g_l$ from different groups $g_k, g_l$. The final loss is defined with a tradeoff constant $\lambda \in [0,1]$ as follows:
\begin{align*}
    \mathcal{L}_{\textit{TD}} &= (1 - \lambda) \mathcal{L}_{Q} + \lambda \mathcal{L}_{\textit{aux}}, \\
\mathcal{L} &= \mathcal{L}_{\textit{TD}} + \mathcal{L}_{\textit{SD}}.
\end{align*}

\section{Experimental Setup} \label{experimental_setup}

\paragraph{Problem Setting Instantiation} We evaluate our approach in the StarCraft Multi-Agent Challenge (SMAC) \citep{samvelyan19smac} to keep our experiments comparable with the previous works.
We train $3-5$ agents with different matching algorithms in the customized scenarios introduced by \citet{ye2024mutual}. The scenario names correspond to specific unit types; for example, SZ represents Stalkers and Zealots, CSZ includes Colossi, Stalkers, and Zealots, and MMM refers to Marines, Marauders, and Medivacs. In evaluation, we use $6-8$ agents with varying leader-follower composition. We report the mean and standard deviation of the winning rate for five random seeds.

\paragraph{Implementation Details} We do not tune the default hyperparameters in REFIL to ensure a fair comparison with baseline approaches (e.g., $\lambda = 0.5$). 

In practice, we use the modified hypernetwork only to generate the weight of the first layer of the mixing networks in Equations \ref{eq2} and \ref{eq3}. Additionally, we use max pooling to generate group states in Figure \ref{fig:hyper_net}. We use four heads for the attention module in the agent's utility network and unify the attention scores with max pooling across different heads. For augmenting hidden states in Figure \ref{fig:utility_net}, we concatenate a one-hot vector of random mapping of group number to a set of size four (i.e., the maximum number of supported groups in our experiment). 
During training, we set  $|\mathcal{L}|=2$
and enforce a team size balancing strategy across all matching algorithms, ensuring equal or nearly equal team sizes among leaders. This strategy remains valid during evaluation, where the number of leaders 
$|\mathcal{L}|$ varies within the range $[2,4]$. Furthermore, in both training and evaluation, the first $|\mathcal{L}|$ agents are designated as leaders.

\section{Results and Discussion} \label{results_section}

\begin{table}[ht]
\renewcommand{\arraystretch}{1.3} % Increases row spacing by 1.5 times
\caption[]{Winning rate on SMAC after $2M$ timesteps}
  \label{table-results1}
  \centering
\begin{tabular}{cccccc}
\hline \multirow{2}{*}{ Tasks } & \multirow{2}{*}{ Algorithms } & Training & \multicolumn{3}{c}{ Evaluation } \\
\cline { 4 - 6 } & & $3-5$ & 6 & 7 & 8 \\
 \hline \multirow{7}{*}{SZ} & MIPI  & $0.659 \pm 0.02$ & $0.453 \pm 0.08$ & $0.404 \pm 0.062$ & 0.276 $\pm 0.076$ \\
\cline { 2 - 6 } & REFIL  & $\mathbf{0.674}$ $\pm 0.038$ & $0.441 \pm 0.103$ & $0.352 \pm 0.078$ & $0.236 \pm 0.103$ \\
\cline { 2 - 6 } & AQMIX  & $0.528 \pm 0.044$ & $0.343 \pm 0.105$ & $0.291 \pm 0.084$ & $0.182 \pm 0.058$ \\
\cline { 2 - 6 } & CollaQ  & 0.588 $\pm 0.03$ & 0.366 $\pm 0.086$ & $0.314 \pm 0.076$ & $0.198 \pm 0.097$ \\
\cline { 2 - 6 } & MAPPO  & $0.256 \pm 0.01$ & $0.129 \pm 0.019$ & $0.148 \pm 0.031$ & $0.036 \pm 0.015$ \\
\cline { 2 - 6 } & OOM*(ours) & 0.624 $\pm 0.010$ & $\mathbf{0.531} \pm 0.037$ & $\mathbf{0.442} \pm 0.044$ & $\mathbf{0.364} \pm 0.019$ \\ %stable
  \cline { 2 - 6 } & SOM*(ours) & 0.639 $\pm 0.007$ & $0.475 \pm 0.049$ & $0.412 \pm 0.041$ & $0.308 \pm 0.022$ \\ %nonstable

\hline \multirow{7}{*}{CSZ} & MIPI  & $0.548 \pm 0.032$ & $0.42 \pm 0.102$ & $0.297 \pm 0.112$ & $0.261 \pm 0.09$ \\
\cline { 2 - 6 } & REFIL & $\mathbf{0.568} \pm 0.027$ & $0.348 \pm 0.057$ & $0.229 \pm 0.053$ & $0.164 \pm 0.06$ \\
\cline { 2 - 6 } & AQMIX & $0.509 \pm 0.054$ & $0.323 \pm 0.096$ & $0.216 \pm 0.101$ & $0.152 \pm 0.071$ \\
\cline { 2 - 6 } & CollaQ & $0.459 \pm 0.061$ & $0.362 \pm 0.13$ & $0.267 \pm 0.099$ & $0.231 \pm 0.095$ \\
\cline { 2 - 6 } & MAPPO & $0.248 \pm 0.037$ & $0.12 \pm 0.029$ & $0.06 \pm 0.028$ & $0.054 \pm 0.013$ \\
\cline { 2 - 6 } & OOM*(ours) & $0.534\pm 0.007$ & $0.399 \pm 0.017$ & $\mathbf{0.317} \pm 0.005$ & $\mathbf{0.329} \pm 0.025$ \\ %stable
  \cline { 2 - 6 } & SOM*(ours) & $0.539 \pm 0.010$ & $\mathbf{0.486} \pm 0.092$ & $0.283 \pm 0.007$ & $0.181 \pm 0.012$ \\ %nonstable

\hline \multirow{7}{*}{MMM} & MIPI & $0.548 \pm 0.023$ & $0.495 \pm 0.054$ & $\mathbf{0.447} \pm 0.041$ & $\mathbf{0.467} \pm 0.067$ \\
\cline { 2 - 6 } & REFIL & $\mathbf{0.605} \pm 0.057$ & $0.437 \pm 0.118$ & $0.329 \pm 0.171$ & $0.224 \pm 0.163$ \\
\cline { 2 - 6 } & AQMIX & $0.501 \pm 0.036$ & $0.447 \pm 0.043$ & $0.344 \pm 0.071$ & $0.251 \pm 0.089$ \\
\cline { 2 - 6 } & CollaQ & $0.589 \pm 0.027$ & $\mathbf{0.513} \pm 0.07$ & $0.423 \pm 0.026$ & $0.286 \pm 0.083$ \\
\cline { 2 - 6 } & MAPPO & $0.289 \pm 0.097$ & $0.32 \pm 0.102$ & $0.25 \pm 0.063$ & $0.275 \pm 0.098$ \\
\cline { 2 - 6 } & OOM*(ours) & $0.586 \pm 0.016$ & $0.443 \pm 0.045$ & $0.394 \pm 0.021$ & $0.332 \pm 0.093$ \\ %stable
  \cline { 2 - 6 } & SOM*(ours) &  $0.498 \pm 0.009$ & $0.318 \pm 0.017$ & $0.219 \pm 0.048$ & $0.165 \pm 0.035$ \\ %nonstable
  \hline
\end{tabular}
\end{table}

Table \ref{table-results1} presents the performance of the best compositions of SOM and OOM compared to prior methods. The results for MIPI, REFIL, CollaQ, AQMIX \citep{iqbal2020ai}, and MAPPO \citep{yu2022surprising} are adopted from a similar experimental setup by \citet{ye2024mutual}. Detailed performance breakdowns for each composition are provided in Tables \ref{table-results-extended_sz}, \ref{table-results-extended_csz}, \ref{table-results-extended_mmm}.
In this work, the added components are specifically designed to operate in tandem with the matching algorithms, forming a tightly coupled framework. As such, ablation of individual modules (e.g., retaining the encoder-decoder structure while removing matching) is not informative. We instead use REFIL as a meaningful ablation-style baseline, which lacks both the matching logic and architectural components, allowing us to evaluate the effect of their integration.
Our findings indicate that in \textbf{6 out of 9} evaluation scenarios, the best compositions in our methods outperform the baselines, while their performance remains comparable in training scenarios. Notably, in \textbf{26 out of 27} evaluation compositions OOM consistently outperforms SOM.
We hypothesize that the superior generalization capability of OOM stems from its stability property. Specifically, by ensuring that no agent has an incentive to deviate from the final matching, OOM may discourage inefficient policy switching caused by frequent group changes. 

A promising direction for future work is to investigate the performance of SOM and OOM in the context of \textit{distracted attention issue}. This phenomenon suggests that increasing agents’ sight range leads attention mechanisms such as REFIL’s to focus on irrelevant context, thereby degrading team performance \citep{shao2023complementary}. Thus, future work can study whether OOM—which uses only the relative order of scores—is more robust to such distractions compared to SOM, which directly relies on attention scores.
Future research can also further investigate the comparative performance of OOM and SOM by incorporating a more robust attention mechanism, such as differential attention \citep{ye2024differential}. Additionally, developing a principled approach for tuning $|\mathcal{L}|$ and selecting leaders, inspired by methods proposed in \citep{shao2022self}, remains an open direction for future work.

\section{Conclusion}

In this work, we introduced a framework to study algorithmic choices for the dynamic grouping of agents. In particular, we investigated two bilateral matching methods and empirically concluded that the stability of the matching algorithm results in teams that show better generalization in unseen tasks.

\subsubsection*{Acknowledgments}
\label{sec:ack}
Computations were performed on the Rouge supercomputer at the SciNet HPC Consortium. SciNet is funded by Innovation, Science and Economic Development Canada; the Digital Research Alliance of Canada; the Ontario Research Fund: Research Excellence; and the University of Toronto.

%%%%%%%%%%%%%%%%%%%%%%%%%%%%%%%%%%%%%%%%%%%%%%%%%%%%%%%%%%%%%%%%
%% Appendices
%%%%%%%%%%%%%%%%%%%%%%%%%%%%%%%%%%%%%%%%%%%%%%%%%%%%%%%%%%%%%%%%
\appendix

\section{Algorithms}
\label{sec:appendix1}

\begin{algorithm}
\caption{Order Oriented Matching}\label{algo:stable_matching}
\begin{algorithmic}[1]
\State Initialize all leaders and followers to be free.
\State Determine the number of spots on each leader's team. 
% \Comment{Total number of spots are equal to number followers}
\While{there is a leader $l$ who has not proposed to every follower and has an empty spot in its team}
    \State Choose such a leader $l$
    \State Let $f$ be the highest-ranked follower in $l$'s preference list to whom $l$ has not yet proposed
    \If{$f$ is on no team}
        \State Add $f$ to the team of $l$
    \Else
        \State Suppose $f$ be on team of $l'$
        \If{$f$ prefers $l'$ to $l$}
            \State No follower will be recruited in $l$'s team
        \Else
            \State $f$ prefers $l$ to $l'$
            \State $f$ will be added to the team of $l$, $l'$ will have one more empty spot in its team
        \EndIf
    \EndIf
\EndWhile
\end{algorithmic}
\end{algorithm}

\begin{algorithm}
\caption{Score Oriented Matching} \label{alg:best_free_leader}
\begin{algorithmic}[1]
\State Let $S$ be a score matrix of size $|\mathcal{A}| \times |\mathcal{A}|$ and $c(l)$ denote team capacity for each leader $l \in L$

\State Initialize an empty team list $\mathcal{G}(l) \gets \emptyset$ for each leader $l \in \mathcal{L}$
\For{each follower $f \in \mathcal{F}$}
    \State Initialize $\text{best\_score} \gets -\infty$
    \State Initialize $\text{best\_leader} \gets \text{None}$
    \For{each leader $l \in L$}
        \If{$|\mathcal{G}(l)| < c(l)$} \Comment{Check if leader $l$ has capacity}
            \State Compute $\text{mutual\_score} \gets S[f][l] + S[l][f]$
            \If{$\text{mutual\_score} > \text{best\_score}$}
                \State $\text{best\_score} \gets \text{mutual\_score}$
                \State $\text{best\_leader} \gets l$
            \EndIf
        \EndIf
    \EndFor
    \If{$\text{best\_leader} \neq \text{None}$}

        \State Add $f$ to $\mathcal{G}(\text{best\_leader})$
    \EndIf
\EndFor

\end{algorithmic}
\end{algorithm}

\newpage
\section{Extended Results}

\begin{table}[ht]
\caption{Winning rate of OOM vs SOM under different compositions in SZ scenario}
  \label{table-results-extended_sz}
  \centering
\renewcommand{\arraystretch}{1.2} % Increases row spacing by 1.5 times
\begin{tabular}{cccccc}
\hline \multirow{2}{*}{ Algorithms } & \multirow{2}{*}{ Leaders } & Training & \multicolumn{3}{c}{ Evaluation } \\
\cline { 4 - 6 } & & $3-5$ & 6 & 7 & 8 \\
 \hline \multirow{3}{*}{OOM} & 2 & $0.624 \pm 0.010$ & $\mathbf{0.531} \pm 0.037$ & $\mathbf{0.442} \pm 0.044$ & $0.329 \pm 0.006$ \\
 \cline { 2 - 6 } & 3 & -  & $0.478 \pm 0.037$ & $0.433 \pm 0.050$ & $0.345 \pm 0.004$ \\
  \cline { 2 - 6 } & 4 & -  & - & - & $\mathbf{0.364} \pm 0.019$ \\
\hline
\multirow{3}{*}{SOM} & 2 & $\mathbf{0.639} \pm 0.007$ & $0.490 \pm 0.038$ & $0.367 \pm 0.039$ & $0.300 \pm 0.018$ \\
 \cline { 2 - 6 } & 3 & -  & $0.475 \pm 0.049$ & $0.412 \pm 0.041$ & $0.308 \pm 0.022$ \\
  \cline { 2 - 6 } & 4 & -  & - & - & $0.300 \pm 0.013$ \\
\end{tabular}
\end{table}

\begin{table}[ht]
\caption{Winning rate of OOM vs SOM under different compositions in CSZ scenario}
  \label{table-results-extended_csz}
  \centering
\renewcommand{\arraystretch}{1.2} % Increases row spacing by 1.5 times
\begin{tabular}{cccccc}
\hline \multirow{2}{*}{ Algorithms } & \multirow{2}{*}{ Leaders } & Training & \multicolumn{3}{c}{ Evaluation } \\
\cline { 4 - 6 } & & $3-5$ & 6 & 7 & 8 \\
 \hline \multirow{3}{*}{OOM} & 2 & $0.534 \pm 0.007$ & $0.399 \pm 0.017$ & $\mathbf{0.317} \pm 0.005$ & $0.297 \pm 0.015$ \\
 \cline { 2 - 6 } & 3 & -  & $0.390 \pm 0.020$ & $0.315 \pm 0.017$ & $\mathbf{0.329} \pm 0.025$ \\
  \cline { 2 - 6 } & 4 & -  & - & - & $0.321 \pm 0.023$ \\
\hline
\multirow{3}{*}{SOM} & 2 & $\mathbf{0.539} \pm 0.010$ & $0.387 \pm 0.073$ & $0.267 \pm 0.010$ & $0.173 \pm 0.017$ \\
 \cline { 2 - 6 } & 3 & -  & $\mathbf{0.486} \pm 0.092$ & $0.283 \pm 0.007$ & $0.181 \pm 0.012$ \\
  \cline { 2 - 6 } & 4 & -  & - & - & $0.173 \pm 0.010$ \\
\end{tabular}
\end{table}

\begin{table}[ht]
\caption{Winning rate of OOM vs SOM under different compositions in MMM scenario}
  \label{table-results-extended_mmm}
  \centering
\renewcommand{\arraystretch}{1.2} % Increases row spacing by 1.5 times
\begin{tabular}{cccccc}
\hline \multirow{2}{*}{ Algorithms } & \multirow{2}{*}{ Leaders } & Training & \multicolumn{3}{c}{ Evaluation } \\
\cline { 4 - 6 } & & $3-5$ & 6 & 7 & 8 \\
 \hline \multirow{3}{*}{OOM} & 2 & $\mathbf{0.586} \pm 0.016$ & $\mathbf{0.443} \pm 0.045$ & $\mathbf{0.394} \pm 0.021$ & $\mathbf{0.332} \pm 0.093$ \\
 \cline { 2 - 6 } & 3 & -  & $0.440 \pm 0.040$ & $0.360 \pm 0.031$ & $0.301 \pm 0.091$ \\
  \cline { 2 - 6 } & 4 & -  & - & - & $0.278 \pm 0.090$ \\
\hline
\multirow{3}{*}{SOM} & 2 & $0.498 \pm 0.009$ & $0.318 \pm 0.017$ & $0.219 \pm 0.048$ & $0.165 \pm 0.035$ \\
 \cline { 2 - 6 } & 3 & -  & $0.305 \pm 0.017$ & $0.199 \pm 0.039$ & $0.159 \pm 0.028$ \\
  \cline { 2 - 6 } & 4 & -  & - & - & $0.140 \pm 0.035$ \\
\end{tabular}
\end{table}

%%%%%%%%%%%%%%%%%%%%%%%%%%%%%%%%%%%%%%%%%%%%%%%%%%%%%%%%%%%%%%%%
%% NOTE: THIS MARKS THE END OF THE "MAIN TEXT"
%%%%%%%%%%%%%%%%%%%%%%%%%%%%%%%%%%%%%%%%%%%%%%%%%%%%%%%%%%%%%%%%

%%%%%%%%%%%%%%%%%%%%%%%%%%%%%%%%%%%%%%%%%%%%%%%%%%%%%%%%%%%%%%%%
%% Bibliography
%%%%%%%%%%%%%%%%%%%%%%%%%%%%%%%%%%%%%%%%%%%%%%%%%%%%%%%%%%%%%%%%
\newpage
\bibliography{main}
\bibliographystyle{rlj}

%%%%%%%%%%%%%%%%%%%%%%%%%%%%%%%%%%%%%%%%%%%%%%%%%%%%%%%%%%%%%%%%
% AUTHOR: If your paper has no supplementary materials, you may 
%         comment out the line below, which creates the title for
%         the supplementary materials.
%%%%%%%%%%%%%%%%%%%%%%%%%%%%%%%%%%%%%%%%%%%%%%%%%%%%%%%%%%%%%%%%
% \beginSupplementaryMaterials

% Content that appears after the references are not part of the ``main text,'' have no page limits, are not necessarily reviewed, and should not contain any claims or material central to the paper. 
% %
% If your paper includes supplementary materials, use the \begin{center}
%     {\tt {\textbackslash}beginSupplementaryMaterials} 
% \end{center}
% command as in this example, which produces the title and disclaimer above. 
% %
% If your paper does not include supplementary materials, this command can be removed or commented out.

\end{document}